# A novel method for surface coverage spectroscopy with atomic force microscope: theory, modeling and experimental results for cylindrical nanostructures

*Francesca Bottacchi,*[1] *Stefano Bottacchi*[2] *and Thomas D. Anthopoulos*[1,3]

[1] Department of Physics and Centre for Plastic Electronics, Blackett Laboratory, Imperial College London, London, SW7 2BW, United Kingdom
[2] Independent Research Consultant, via don C. Gnocchi 33, 20148, Milan, Italy
[3] King Abdullah University of Science and Technology (KAUST), Division of Physical Sciences and Engineering and KAUST Solar Centre, Thuwal 23955-6900, Saudi Arabia

## Abstract

A novel method for measuring the surface coverage of randomly distributed cylindrical nanoparticles (nanorods, nanowires *etc*.) using conventional atomic force microscopy (AFM), is presented. The method offers several advantages over existing techniques such as particle-beam and x-ray diffraction spectroscopy. These include, sub-/nanometer vertical and lateral resolution, non-destructive interaction with the sample's surface allowing repeated measurements, user-friendly setup and ambient operating conditions. The method relies on the use of a statistical model to describe the variations of the nanoparticles aggregates height as a function of *x*-*y* position on the sample's surface measured by AFM. To verify the validity of the method we studied two types of randomly oriented networks of carbon nanotubes (CNTs) and silver nanowires (Ag NWs) both processed from solution-phase. Experimental results are found to be in excellent agreement with model's predictions whilst analysis of the measured surface height density, together with the nanoparticle's diameter statistical distribution, allow the extraction of the coverage coefficients for all detected nanoparticle aggregates, as well as for the total surface coverage. The method can be seen as a new powerful tool for the quantitative surface coverage analysis of arbitrary nanoscale systems.

## 1. Introduction

Quantitative evaluation of surface coverage at the nanoscale[1] is a fundamental requirement for many modern surface science applications. The characterization of chemical vapor deposited films[2] has usually been accomplished by scanning electron microscopy (SEM) or transmission electron microscopy (TEM).[3] However, these techniques are expensive, destructive, rely on vacuum, often suffers from surface charging, and at their best, provide semi-quantitative information about surface coverage. Hence, there is a need for an in-situ technique with high spatial resolution to evaluate the surface coverage of adsorbed species and nanostructures under atmospheric conditions. Recently, optical methods such as sum frequency generation,[4] infrared spectroscopy,[5,6] second harmonic generation[7–9] or fluorescence-based techniques[10,11] introduced significant advantages over conventional particle-beam and x-ray diffraction spectroscopy, due to their surface sensitivity, spatial resolution, non-destructive interaction with soft samples and ambient operating conditions.

Another class of techniques that possess all desired characteristics is the so-called scanning probe techniques (SPM). The latter includes atomic force microscopy (AFM),[12,13] magnetic force microscopy (MFM),[14–17] scanning tunneling microscopy (STM),[18–21] kelvin force microscopy (KFM),[22–26] and scanning near-field optical microscopy (SNOM).[27–30] SPM techniques have so far been used for morphological, electrical, optical and magnetic characterization of the sample's surface, with nanoscale accuracy. However, to the best of our knowledge, none of these methods have ever been used to quantify the surface coverage of a certain nanomaterial deposited on a solid substrate.

Here, we present a novel method that can be used to quantify the surface coverage of cylindrical nanostructures like carbon nanotubes (CNTs) and silver nanowires (Ag NWs), deposited from solution onto different substrates, by exploiting topographic information acquired via standard tapping-mode AFM. We develop a statistical model of the height density of the cylindrical nanostructure as a function of the diameter distribution and apply it to the experimental results to extract the coverage coefficients of all measured height configurations deposited onto the substrate. The validity of the method is demonstrated by extracting the coverage spectroscopic coefficients for random networks of CNTs and Ag NWs deposited on $SiO_2$ and glass substrates.

## 2. Method: Theory of the Coverage Spectroscopy[1]

Aim of this work is the evaluation of the coverage distribution of randomly distributed nanoparticles/nanostructures over a substrate, through high resolution height measurements performed by AFM. The roughness of the substrate and the random distribution of the nanostructures suggest using a statistical model to describe the variation of the height versus the position on the sample surface. The height density of the randomly distributed nanostructures over the substrate, together with the diameter density function of the selected nanostructure, allows the calculation of the coverage spectroscopy of the measured sample. For clarity, some of the equations mentioned here have been listed in the Appendix.

### 2.1 The Delta Model

Let $\underline{u}$ and $\underline{v}$ be two random variables representing respectively the height of the substrate and the height of the cylindrical nanostructures distribution, i.e. CNT, each measured independently with respect to a reference plane. The Delta model approximation described here is based on the following three assumptions:

i. The diameter is a deterministic variable with constant value *d*.

ii. The height density function $f_{\underline{v}}^{(1)}(z)$ of the single-layer nanostructure is represented by the Delta distribution centered at its diameter, as shown in (A.1).

iii. The height $\underline{v}^{(k)}$ of the superposition of *k* nanostructures is given by the sum of *k* independent random variables $\underline{v}_j$ and the density function of the total height is given by $k-1$ times the self-convolution [30] of the individual height density, as presented in (A.2).

Because of the properties of the Delta distribution, the density function of the superposition of *k* nanostructures is the Delta distribution located at the integer multiple *kd* of the diameter. It is important to note that this conclusion is just an application of the average theorem of multiple convolutions: *the mean value of the convolution between multiple density functions coincides with the sum of the mean values of the convolving densities.*[31,32]

According to the *Bayes* theorem[31] of the total probability, the density function of the total height variable is given by the sum of the product of the probability of occurrence of each nanostructure superposition by the density of the corresponding height variable. Let $c_k$ be the probability of occurrence of $k$ pile-up nanostructures. Assuming there is no substrate, the density of the total height is given by the sum in (A.3). The probability of occurrence $c_k$ is the coverage coefficient of the $k$ pile-up nanostructure configuration. Given the population of all cylindrical nanostructures distributed over the measured substrate, the total coverage probability must be unitary (A.4). Since the coverage coefficient $c_0$ corresponds to the uncovered configuration, we conclude that the total coverage $C_N$, resulting from $N$ nanostructures configurations, satisfies (A.5).

### 2.1.1 The Coverage Equation

The height variable of the substrate is modeled with the random variable $\underline{u}$ with density $g_{\underline{u}}(z)$. Then, the substrate height is added to each nanostructure height variable $\underline{v}^{(k)}$, $k = 0,1 \ldots N$. Since $\underline{u}$ and $\underline{v}_j$, $j = 0,1 \ldots k$, $k = 0,1 \ldots N$ constitute a set of mutually independent random variables,[31] the density of the total height variable $\underline{z}$ is given by the convolution of $f_{\underline{v}}(z)$ (A.3) with the substrate density $g_{\underline{u}}(z)$ (A.6). Substituting (A.3) into (A.6), we conclude that the density function $f_{\underline{z}}(z)$ of the total height of the random superposition of nanostructures over the substrate with known height density function, satisfies the *coverage equation* (1) in the Delta model approximation:

$$f_{\underline{z}}(z) = \sum_{k=0}^{N} c_k g_{\underline{u}}(z - kd) \tag{1}$$

From (1) we conclude that the coverage equation is given by the weighted superposition of the substrate density function translated at the integer multiples of the deterministic diameter of the specified nanostructure. The weighting factors $c_k$ are the coverage coefficients.

### 2.1.2 The Linear System of Equations

Assuming the substrate has the known height density $g_{\underline{u}}(z)$, the coverage equation (1) allows the determination of all coverage coefficients $c_k$, $k = 0, 1 \ldots N$. To this end, we form the linear system of $N$

+ 1 linearly independent equations in $N + 1$ unknowns, namely the coverage coefficients $c_k$, $k = 0, 1\ldots N$ (A.7). The factors $B_j$ are the samples of the total height density function measured at every integer multiple of the diameter, from the origin.

**Figure 1** illustrates the meaning of the terms in equation (A.7). By knowing the substrate height density and the measured samples $B_j$, we obtain the solution of the coverage spectroscopy from the linear system shown in (A.8) and (A.9). Finally, assuming the substrate has a Gaussian height density with mean $z_s$ and standard deviation $\sigma_s$, the system coefficients $a_{jk}$ in (A.8) assume the simple form of (A.10). The assumption of the Gaussian substrate allows the determination of the coverage coefficients in (A.9) by knowing only four parameters, more specifically the mean substrate height $z_s$, the substrate standard deviation $\sigma_s$, the diameter $d$ of the cylindrical nanostructures and the $N + 1$ samples $B_j$ of the measured height density.

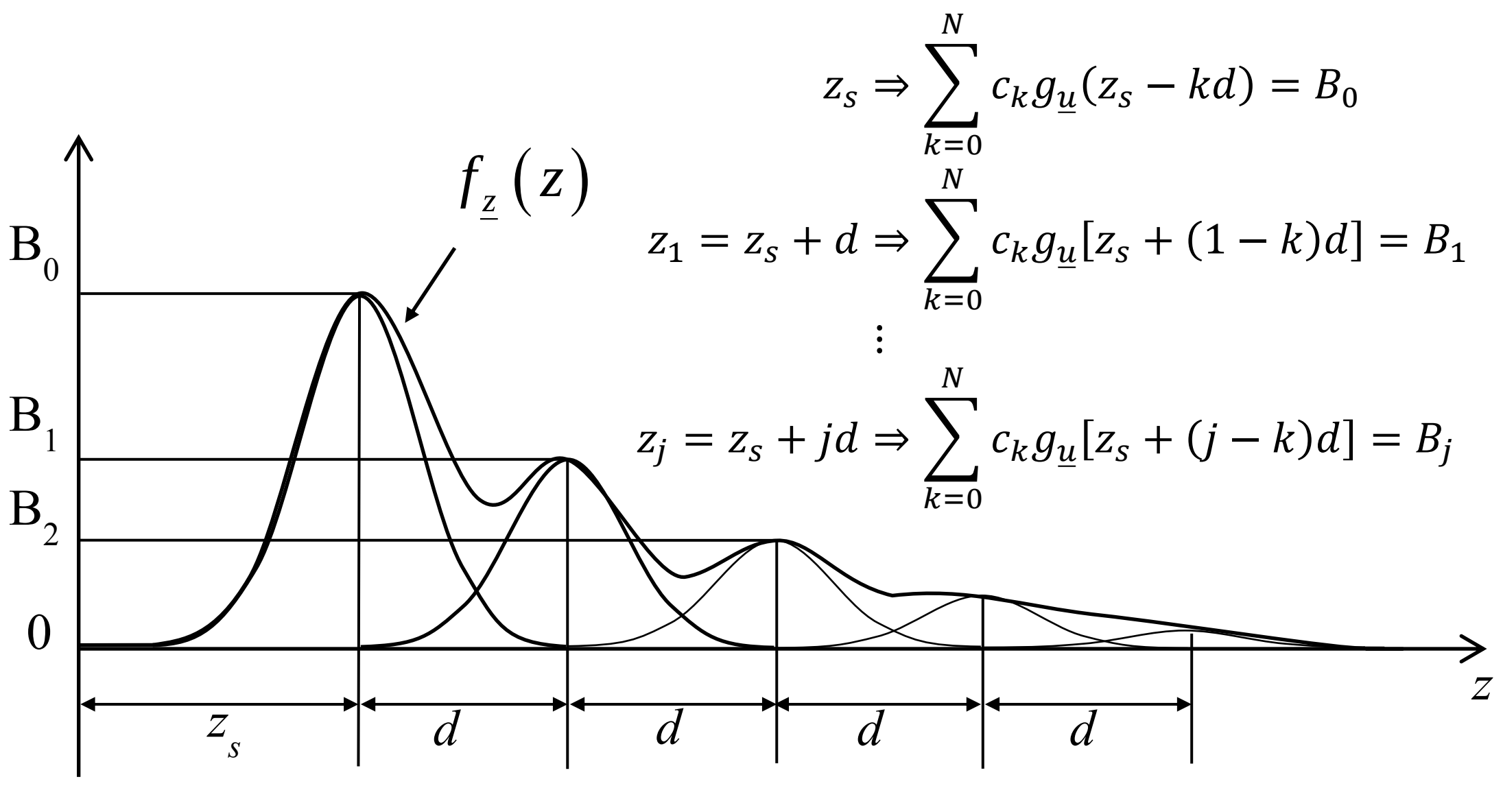

**Figure 1:** Illustration of the terms in the system of equations (A.7) for the solution of the coverage spectroscopy in the Delta model approximation.

### *2.1.3 Limitations of the Delta Model*

Besides its simplicity, the Delta model suffers from some limitations and it must be generalized to account for the experimental evidence:

i. The height density of the deterministic diameter is roughly approximated by the Delta distribution. In fact, it is apparent that to justify the continuous distribution of measured height values, as resulting from the random superposition of cylindrical nanostructures, the height density function cannot be represented by the simple Delta distribution.

ii. Different diameter statistics must be accounted to represent heterogeneous populations of cylindrical nanostructures.

iii. The Delta model requires that the substrate roughness is comparable to the nanostructure diameter, providing satisfactory results only when the Gaussian substrate dominates over the specific height densities of the cylindrical nanostructures.

iv. The substrate is only approximately Gaussian and thus it needs a more general model.

## 2.2 Statistical Height Models

The motivation for the generalized coverage theory is to resolve the limitations of the ideal Delta model, providing a simulation environment more suitable for the experimental evidence. Here, we formulate the height density theory of cylindrical nanostructures with random diameter distributions as depicted in **Figure 2**.

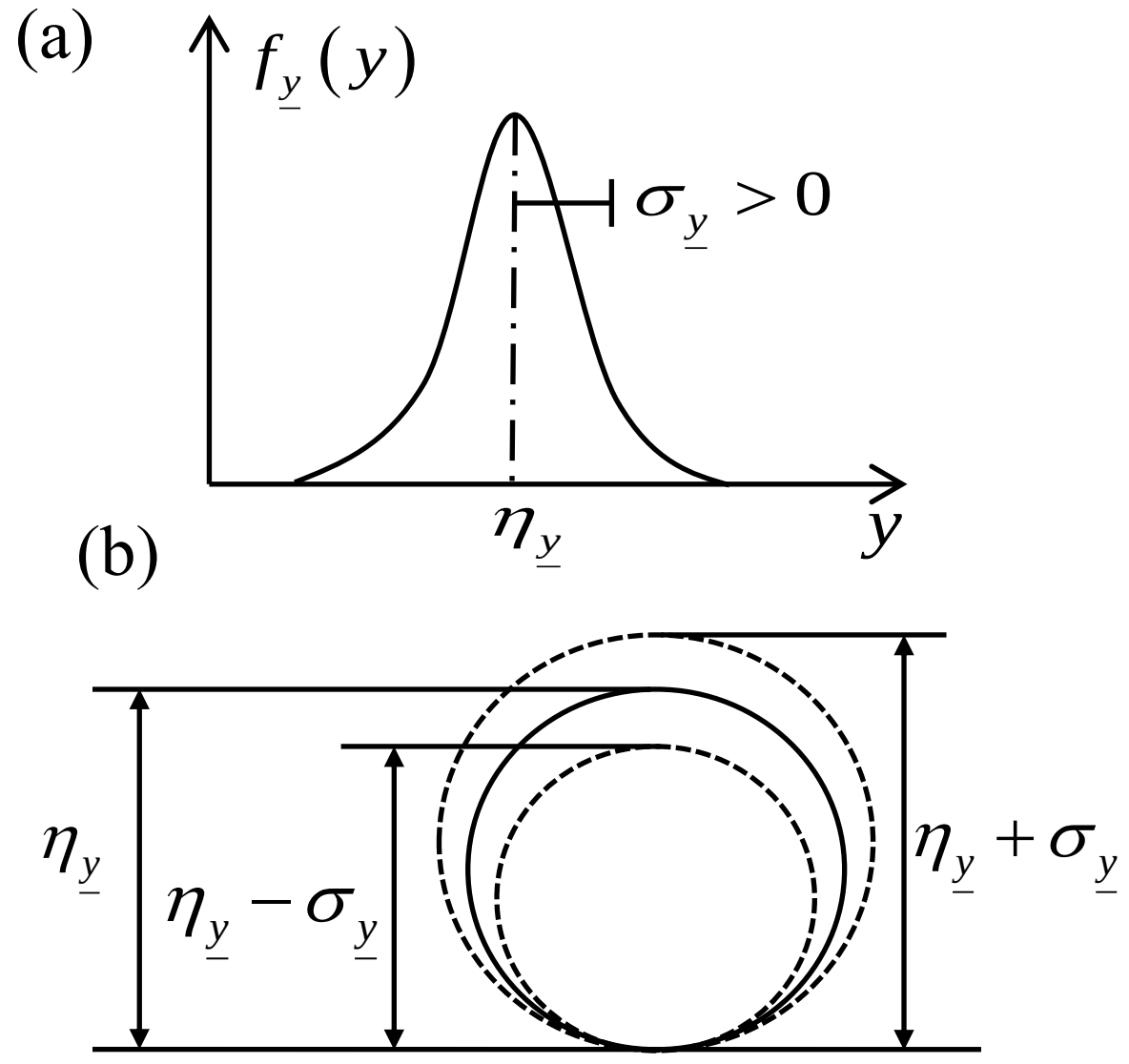

**Figure 2:** Schematics of (a) the diameter density function and (b) of the geometrical representation of the cylindrical cross-section with the expected variation of the diameter.

The statistical model of the height variable $\underline{y}$ of the cylinder with a random diameter requires the derivation of the joint probability density function[31] between the horizontal position $\underline{x}$ of the AFM probe and the random diameter $\underline{y}$ (A.11). For this purpose, we formulate the following assumptions:

1. The random variable $\underline{x}$ is uniformly distributed in the interval $|x| \le y/2$ with the conditional probability density indicated in (A.11).
2. The joint density $f_{\underline{xy}}(x,y)$ (A.12) is given by the product of the conditioned density $f_{x|y}(x)$ with the probability density $f_y(y)$ of the diameter $\underline{y}$.[31]
3. The height density function $f_{\underline{v}}(z)$ of the cylindrical cross-section with random diameter $\underline{y}$ is then obtained integrating the diameter density $f_y(y)$ with the proper surface weight function:

$$f_{\underline{v}}(z) = \int_{I_y(z)} \frac{2z-y}{y\sqrt{yz-z^2}} f_{\underline{y}}(y)\,dy \quad , \quad z \in I_z \tag{2}$$

It is important to remark that for every diameter density, the height density function in (2) is normalized, as indicated in (A.13).

In the following sections, we apply the general equation (2) of the height density function of cylindrical structures to the deterministic, uniform and Gaussian-Harmonic random diameter distributions.

### *2.2.1 Deterministic Diameter*

The deterministic diameter is modelled with the impulsive density located at the diameter value *d*. From the general height density (2), after simple calculations, we obtain the height density of the cylinder with the deterministic diameter:

$$f_{\underline{y}}(y) = \delta(y-d) \quad \Rightarrow \quad f_{\underline{v}}(z) = \begin{cases} \dfrac{1}{d}\dfrac{2z-d}{\sqrt{dz-z^2}} , & \dfrac{d}{2} \le z \le d \\ 0 \quad , & z \le \dfrac{d}{2} \cup z \ge d \end{cases} \tag{3}$$

The mean and the standard deviation of the height variable are reported in (A.14) and (A.15).

**Figure 3** shows the computed height density (3) of the cylindrical nanoparticle with the deterministic diameter $d$ = 80 nm. It is evident from the plot that the density function is zero for $z \le d/2$ = 40 nm and for $z > d$ = 80 nm, while at $z = d$ = 80 nm the density function is singular. Due to the singularity at the diameter height, the density function of the deterministic diameter could be roughly approximated by the Delta distribution.

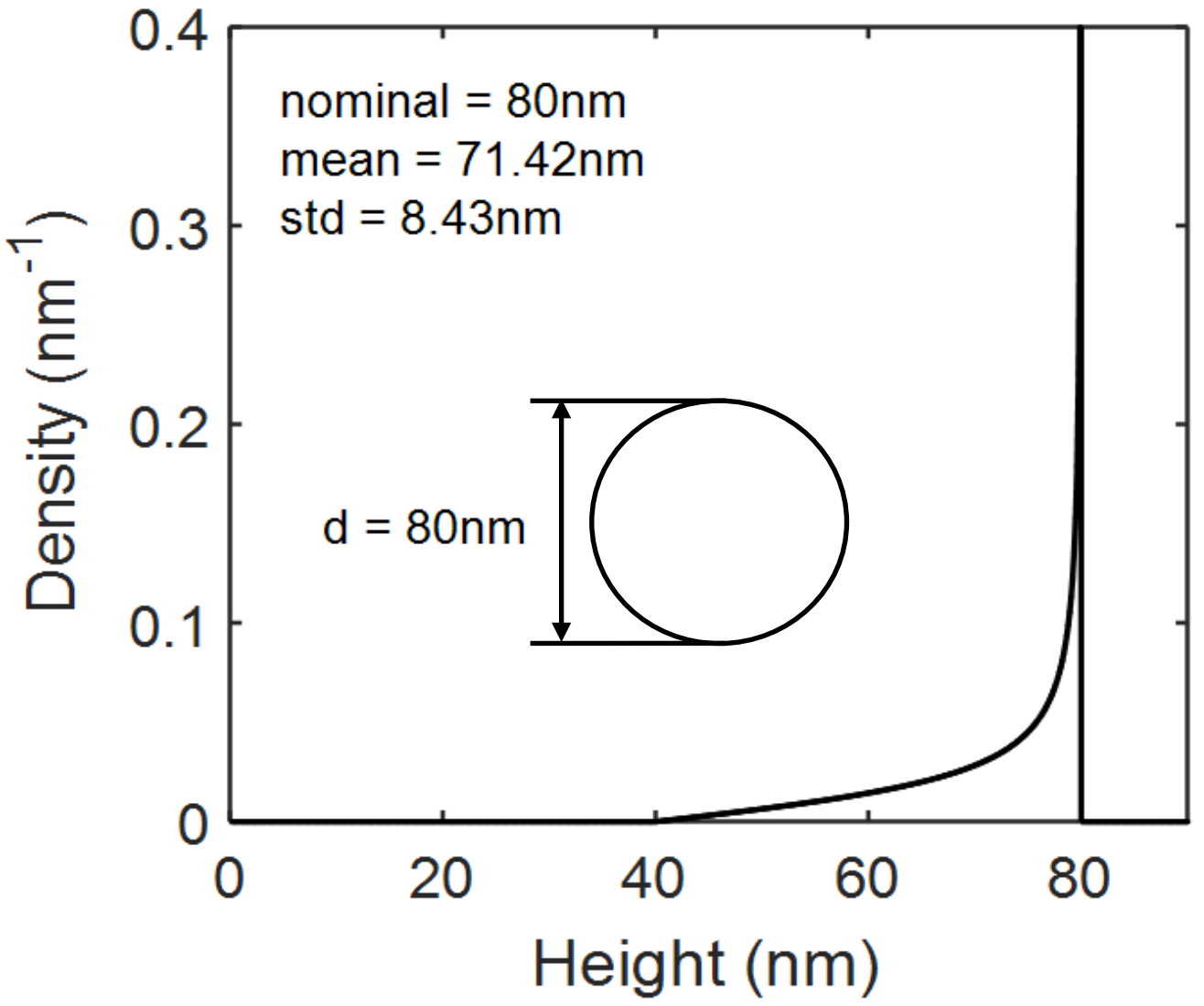

**Figure 3:** Simulation of the height density function of a cylinder with deterministic diameter *d* = 80 nm. The density is identically zero for any height below half-diameter, hence 40 nm in the case shown.

### *2.2.2 Uniform Diameter*

The uniform diameter is modelled with the constant density function centered on the nominal value *d*, with the full width specified by the tolerance range Δ, as indicated in (A.16) and **Figure 4**.

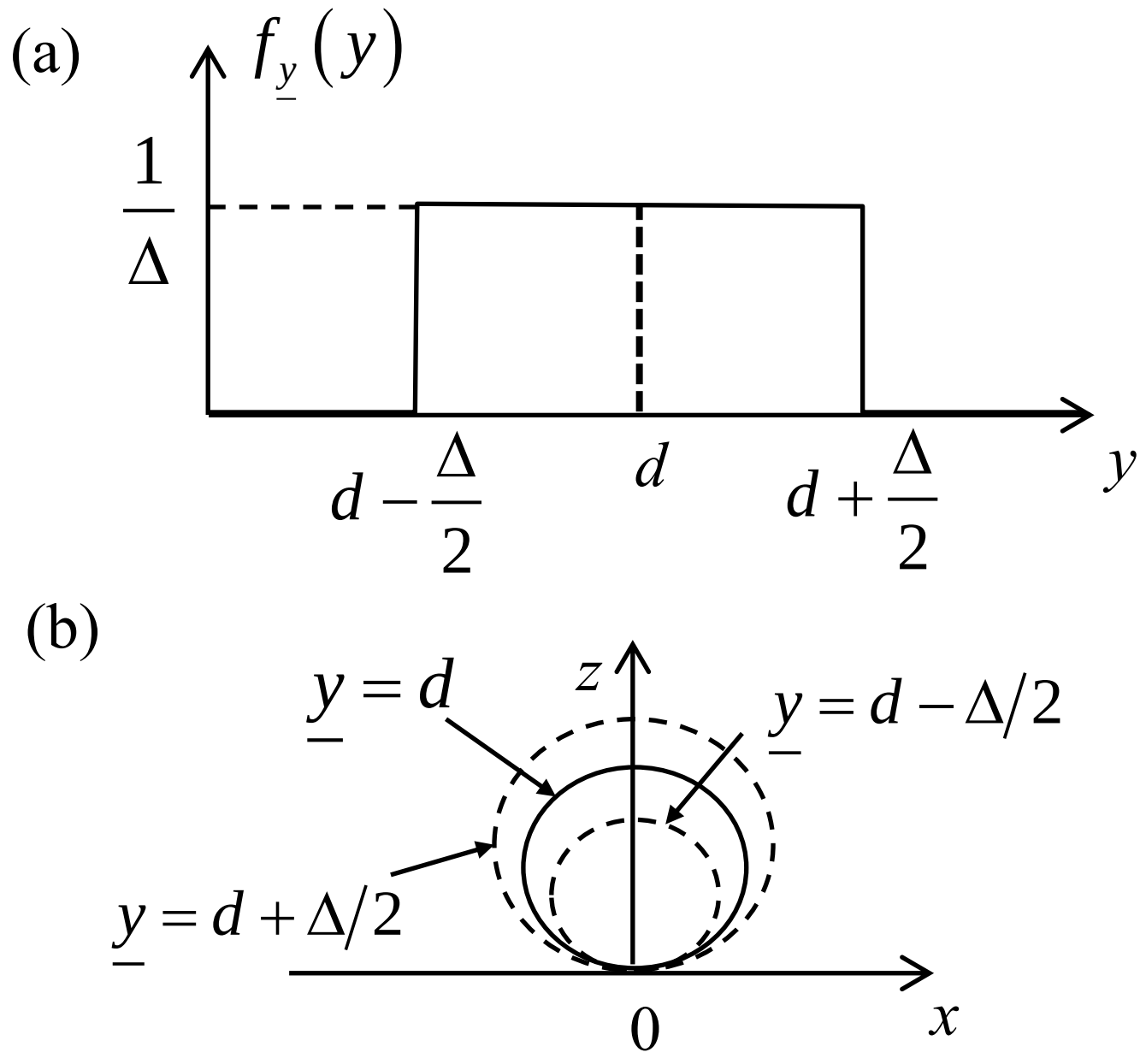

**Figure 4:** Schematics of (a) the uniform diameter density function and of (b) the geometrical representation of the cylindrical cross-section with the expected variation of the diameter.

The height density of the cylinder with a uniform diameter, nominal value $d$ and tolerance $\Delta$, is obtained by substituting (A.16) into the general form (2). After some calculations, we obtain the following equation**:**

$$f_{\underline{v}}(z)=\frac{2}{\Delta}\left\{\begin{array}{l}2\left[\arctan\sqrt{\frac{1}{z}\min\left(2z,d+\frac{\Delta}{2}\right)-1}-\arctan\sqrt{\frac{1}{z}\max\left(z,d-\frac{\Delta}{2}\right)-1}\right]-\\-\left[\sqrt{\frac{1}{z}\min\left(2z,d+\frac{\Delta}{2}\right)-1}-\sqrt{\frac{1}{z}\max\left(z,d-\frac{\Delta}{2}\right)-1}\right]\end{array}\right\} \qquad (4)$$

$$I_z=\left[\frac{d}{2}-\frac{\Delta}{4},d+\frac{\Delta}{2}\right]$$

**Figure 5** shows the simulated plot of the height density function (4) of several cylindrical nanostructures with the same nominal diameter but with different tolerances and uniform distribution.

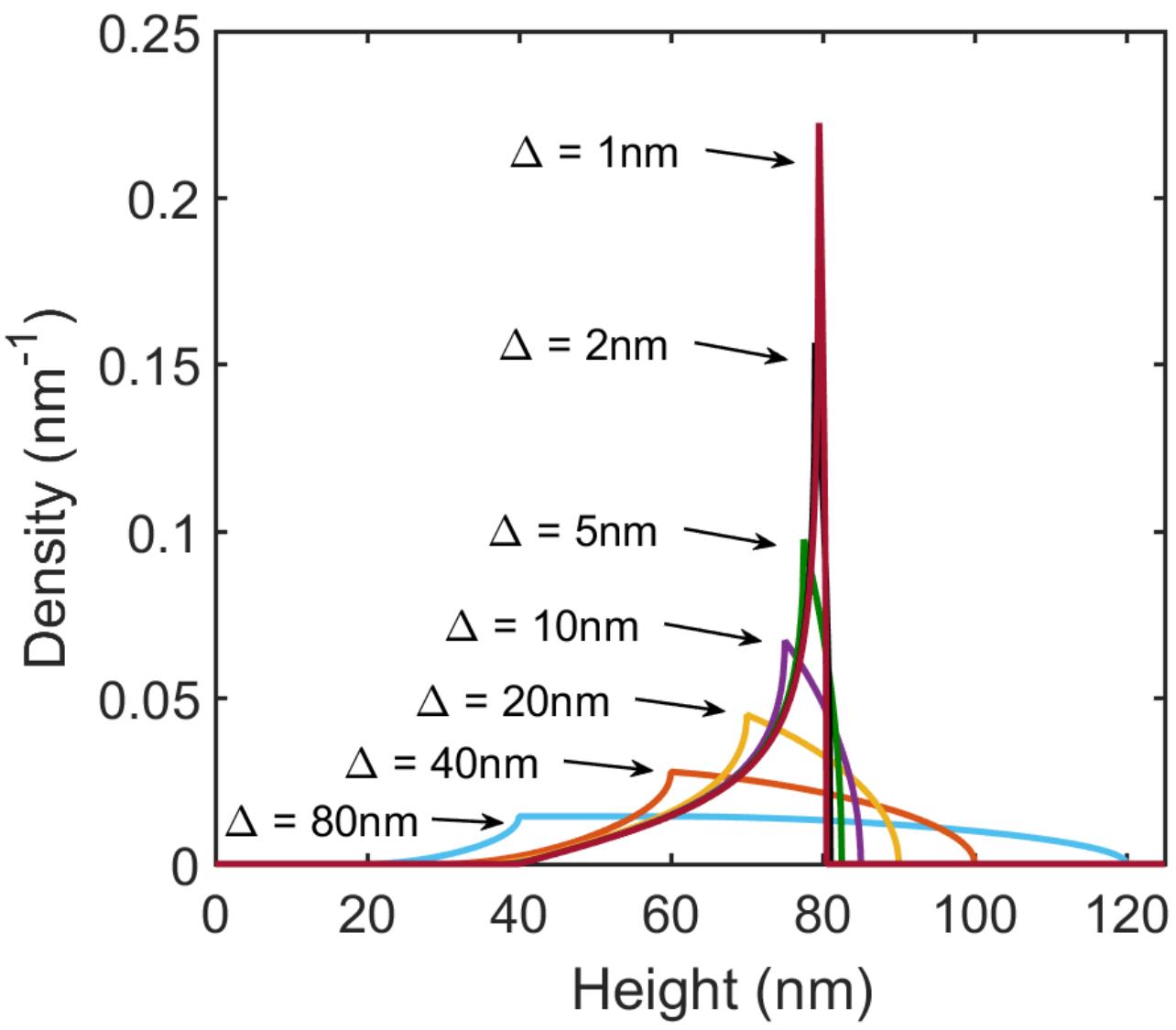

**Figure 5:** Simulation of the height densities of cylinders with uniform random diameters. All curves refer to the same nominal diameter $d = 80$ nm but with different tolerances $\Delta$.

It is evident that by reducing the tolerance, the height density approaches the case of the deterministic diameter shown in **Figure 3**.

### *2.2.3 Gaussian-Harmonic Diameter*

The Gaussian-Harmonic probability density function is a generalization of the Rayleigh probability density where both orthogonal amplitudes $\underline{a}$ and $\underline{b}$ are normal distributed random variables with the same variance $\sigma$ but with non-zero mean $\rho$, as shown in (A.17).

Let us define the nominal diameter as $d = 2\rho\sqrt{2}$ and the random diameter $\underline{y}$ equals to the double of the geometric mean between the normal random variables $\underline{a}$ and $\underline{b}$, according to (A.18). With simple calculations, we conclude that the probability density function of the diameter (2) coincides with the overlapping integral between the circular symmetric domain centered at the origin and the center symmetric Gaussian joint density $f_{\underline{ab}}(\xi,\eta)$ centered at the position $(\rho,\rho)$. From (A.17) and (A.18), we conclude that the density function of the Gaussian-Harmonic diameter is given as:

$$f_{\underline{y}}(y) = \frac{1}{4\sigma^2} y e^{-\frac{y^2+d^2}{8\sigma^2}} I_0\left(\frac{yd}{4\sigma^2}\right) \quad , \quad y \geq 0 \tag{5}$$

When the ratio between the mean value $\rho$ (A.17) and the standard deviation $\sigma$ becomes large, the modified Bessel function of first kind and zero order in the equation (5) is well approximated by the exponential function and the density function of the Gaussian-Harmonic diameter approaches the symmetric Gaussian profile of equation (A.19). In general, the mean and the standard deviation of the Gaussian-Harmonic diameter depend from the variables $\rho$ and $\sigma$ through integral equations that can be solved using numerical methods. The mean is always larger than the peak position and it approaches the peak for very large ratios $d/\sigma$.

The simulations of the height densities generated by the circular cross-section with the Gaussian-Harmonic diameter distribution are shown in **Figure 6**. The curves report the height density with the fixed nominal diameter $d$ = 80nm, versus different tolerances characterized by the parameter $\sigma_d$. It is apparent that when the ratio $d/\sigma_d$ becomes relatively large, the density profile approaches the same highly peaked shape as the height density obtained from the deterministic diameter shown in **Figure 3**, verifying the correctness of the Gaussian-Harmonic model.

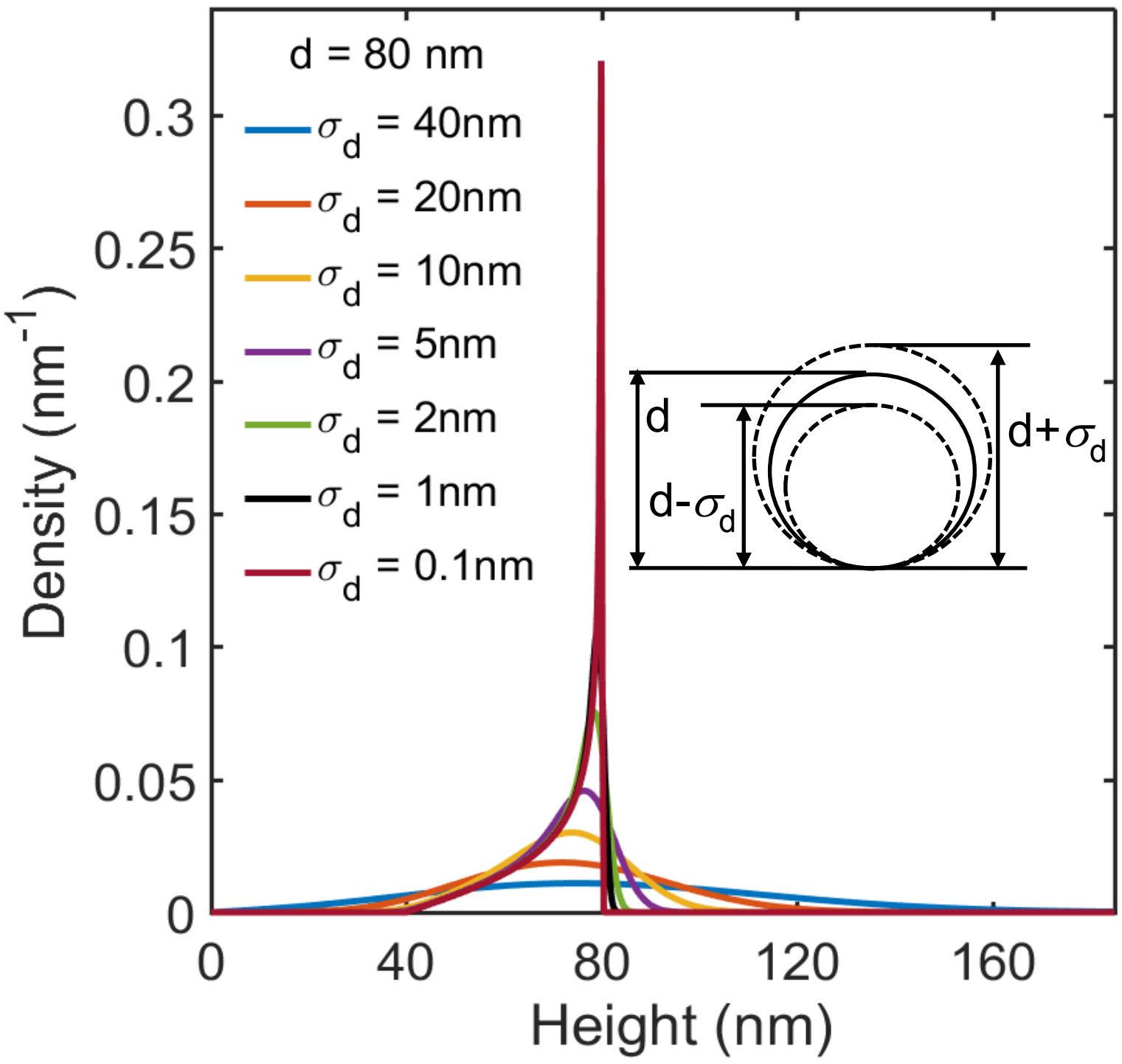

**Figure 6:** Computed plots of the height density for the Gaussian-Harmonic diameter density function at decreasing values of the standard deviation. The height density at relatively large tolerance becomes almost symmetric and it is well approximated by a Gaussian profile. Decreasing the standard deviation of the Gaussian-Harmonic diameter density function, the height density begins peaking, loosing gradually the symmetry and approaching the highly-peaked profile obtained by the Delta distribution of the diameter density. We remark that the Gaussian-Harmonic density approaches the Delta distribution at infinitesimal values of the standard deviation.

## 2.3 The Generalized Coverage Theory

The motivation of the general coverage theory is two-fold:

1. Confirm the successful results we have verified with the Delta model when the measurement conditions were dominated by the Gaussian density of the substrate.
2. Extract the coverage coefficients under general measurement conditions, implementing a realistic statistical height model of both cylindrical nanostructures and substrate.

### *2.3.1 Axioms and Assumptions*

In the following section, we list the assumptions used to develop the general coverage theory. To begin with, we assume that all cylindrical elements have the diameter distributed with the same known density function $f_{\underline{y}}(y)$. In particular, it can be deterministic, uniform, or Gaussian-Harmonic. The height density $f_{\underline{v}}(z)$ of the single circular cross-section is a known function of the diameter statistic. The events {*j-stacked cylinders*} and {*k-stacked cylinders*} are statistically independent, with probabilities of occurrence respectively $c_j$ and $c_k$, known as partial coverage. The total height $\hat{\underline{z}}^{(k)}$ of $k$ stacked cylinders, without the substrate contribution, is a random variable given by the sum of the height $\underline{z}_i$, $i = 1,\ldots k$ of each element of the cylinders aggregate, as reported in (A.20). The heights $\underline{z}_i$ are mutually statistically independent random variables, distributed with the same density function $f_{\underline{v}}(z)$. The conditioned density function $f_{\underline{v}|k}(z)$ of the height variable $\hat{\underline{z}}^{(k)}$ of $k$ stacked cylinders is given by $k$ – 1 times the self-convolution[31] of the individual density $f_{\underline{v}}(z)$, as shown in (A.21). The total height $\underline{z}^{(k)}$ of $k$ stacked cylinders placed upon the substrate is given by the sum of the height $\hat{\underline{z}}^{(k)}$ of the $k$ stacked cylinders in (A.20) with the substrate height $\underline{z}_s$, as indicated in (A.22). The height $\underline{z}_s$ of the substrate and the height $\underline{z}_i$ of each cylinder in the stacking aggregate form a set of mutually statistically independent random variables.

The conditioned density function $f_{\underline{z}|k}(z)$ (A.23) of the total height $\underline{z}^{(k)}$ of $k$ stacked cylinders placed upon the substrate, is given by the convolution[31] of the height density $g_{\underline{u}}(z)$ of the substrate with the conditioned density $f_{\underline{v}|k}(z)$.

a. *Coverage Master Equation in the Physical Domain* – The total height density $f_{\underline{z}}(z)$ of the entire cylinders' population stacked in $N$ different configurations and placed upon the substrate, is given by the linear combination of $N + 1$ conditioned probability density functions from (A.23), each weighted by the probability of the corresponding stacked configuration:

$$f_{\underline{z}}(z) = \sum_{k=0}^{N} c_k \, g(z) * f_{\underline{v}|k}(z) \tag{6}$$

The probability $c_k$ of the event {*k-stacked cylinders*} assumes the meaning of the coverage coefficient for that event, i.e. how many $k$ – *stacked* cylinders are present over the entire cylinders' population. The coverage coefficients $c_k$ must satisfy the normalization condition for the total probability, as shown in (A.24). In particular, the coverage coefficient $c_0$ assumes the meaning of the *uncovered* substrate percentage. Accordingly, the sum of all coverage coefficients between the single layer and the $N$ – *stacked* layer configurations represents the total coverage $C$ of the cylindrical nanoparticles placed over the given substrate, as reported in (A.25). Finally, the conditioned probability density function $f_{\underline{v}|k}(z)$ (A.26) can be conveniently calculated using the convolution theorem of the Fourier integral.[32]

b. *Coverage Master Equation in the Conjugate Domain* – From the coverage master equation (6), we deduce that the total height density $f_{\underline{z}}(z)$ is given by the inverse Fourier transform of the linear combination, through the coverage coefficients $c_k$, of the products between the Fourier transform of the height density of the Gaussian substrate with the $k-th$ power of the Fourier transform of the height density of the circular cross-section,[31] corresponding to the selected diameter statistic:

$$g_{\underline{u}}(z) \stackrel{\Im}{\longleftrightarrow} G_{\underline{u}}(\xi) \quad , \quad f_{\underline{z}}(z) \stackrel{\Im}{\longleftrightarrow} F_{\underline{z}}(\xi) = G_{\underline{u}}(\xi) \sum_{k=0}^{N} c_k F_{\underline{v}}^{k}(\xi) \tag{7}$$

Equation (7) constitutes the coverage master equation in the conjugate domain.

### 2.3.2 System of Coverage Equations

The unknown coefficients $c_0, c_1 \ldots c_N$ of the coverage master equations, either in the form (6) or (7), are the solution of the system of $N + 1$ independent linear equations obtained sampling the measured height density at specified height positions. The positions of the height samples can be chosen arbitrarily; possible choices are the multiples of the nominal diameter or the positions of the mean height at increasing stacking levels. Choosing to sample at integer multiples of the diameter, we obtain the sequence shown in (A.27). Each sample $B_j$ of the measured height density evaluated at $z_j$ satisfies the coverage master equation (6). Providing $N + 1$ height samples of the measured density profile, we obtain the following system of $N + 1$ independent linear equations:

$$\sum_{k=0}^{N} c_k f_{\underline{z}|k}(z_s + jd) = B_j \ , \quad j = 0,1,\ldots,N \tag{8}$$

*2.3.3 Matrix Representation*

The linear system of equations (8) can be easily represented in the matrix form. To this end, we introduce the matrix elements in (A.28), and we define the system matrices in (A.29). Substituting (A.28) and (A.29) into (8), the expected matrix form of the coverage master equation is then obtained:

$$\sum_{k=0}^{N} a_{jk} c_k = B_j, \quad j = 0,1,\ldots,N \quad \Rightarrow \quad \mathbf{Ac} = \mathbf{B} \tag{9}$$

The matrix elements in (A.28) can be conveniently calculated using the convolution theorem [31] of the Fourier transform, as indicated in the following equation:

$$a_{jk} = \mathfrak{I}^{-1}\left[ G_{\underline{u}}\left(\xi\right) F_{\underline{v}}^{k}\left(\xi\right)\right]_{z=z_s+jd}, \quad j,k = 0,1,\ldots,N \tag{10}$$

Finally, the solution of the linear system for the coverage spectroscopy can be obtained by standard numerical methods.

## 3. Results and Discussion

In this section, we discuss the applications of the coverage theory to AFM measurements performed on solution-processed single chirality (7,5) carbon nanotubes deposited on $SiO_2$ substrates and silver nanowires deposited from solution both on glass and $SiO_2$ substrates.

### 3.1 Carbon Nanotubes

**Figure 7** illustrates the topography (a) and the height density (b) of (7,5) CNTs wrapped with PFO deposited onto a $SiO_2$ substrate, measured with AFM in tapping-mode operation. It is evident from the topography the high coverage factor of the CNT networks and the high stacking layer combinations. Although the $SiO_2$ substrate height is expected to be in the order of 1-2 nm with respect to the instrument reference and to have sub-nm roughness, the height density extends considerably close to 10 nm, indicating a large number of CNT stacking layers.

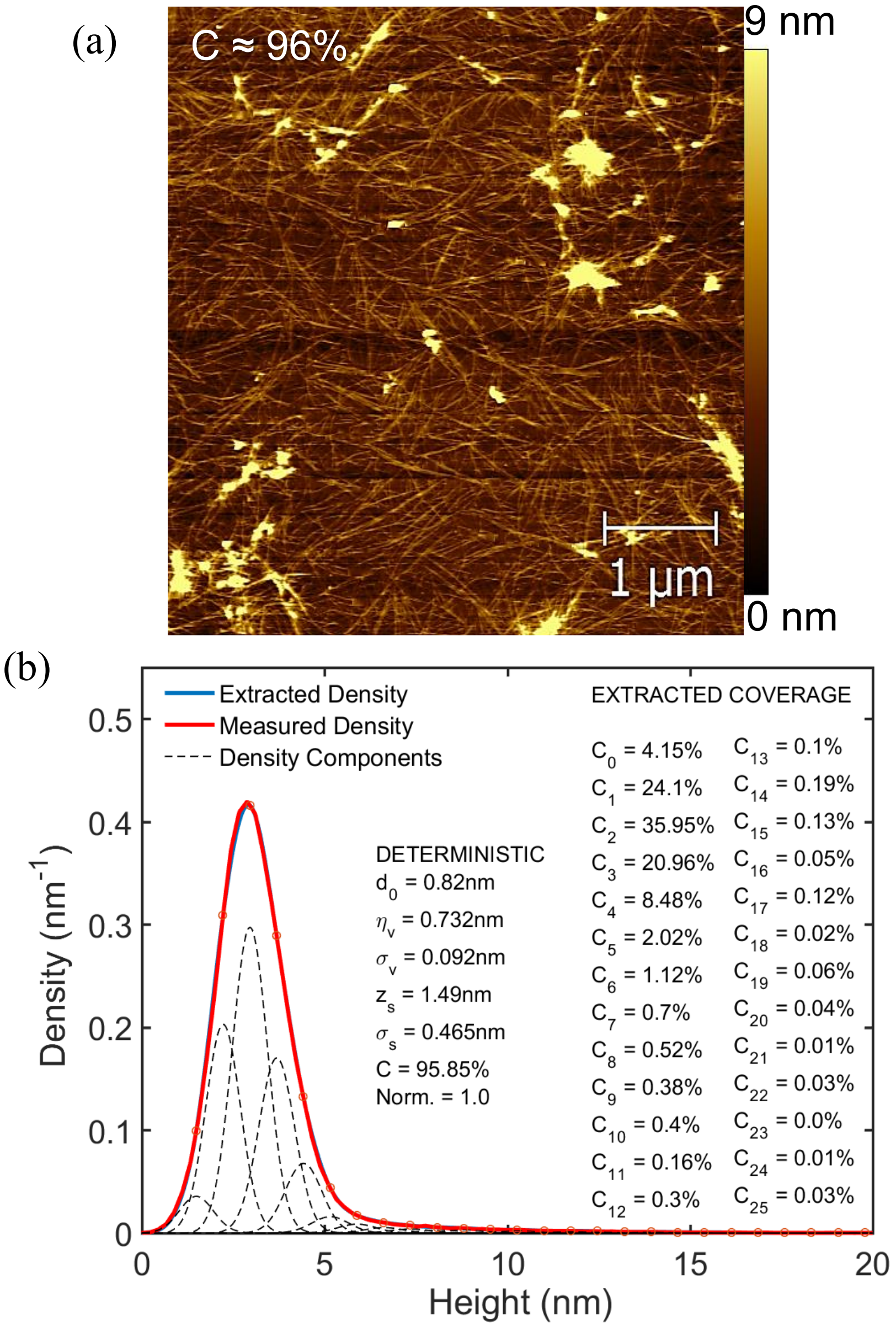

**Figure 7:** (a) Measured topography and (b) height density profile obtained with the AFM operated in tapping-mode on the solution processed PFO/(7,5) CNT network randomly distributed over the $SiO_2$ substrate. The coverage spectroscopy solution assumes the deterministic diameter model and it is shown in the inset of (b). The computed total coverage results $C$ = 95.85%, with dominant dual-layer CNT configurations, resulting into the coverage coefficient $c_2$ = 35.95%.

The measured profile has been captured up to 20 nm of maximum height, thus including residual high-order CNT stacked configurations. The coverage has been extracted assuming the deterministic diameter model of cylindrical (7,5) CNT with nominal diameter $d$ = 0.82 nm.[33] The substrate height has been modelled using the Gaussian density with mean height $z_s$ = 1.49 nm and standard deviation $\sigma_s$ = 0.465 nm. Because of the large value of the maximum height $z_{max}$ = 20 nm considered in this measurement, the number of allowed coverage coefficients become very large as well, counting $N = \text{int}\left[\left(z_{\max} - \overline{z}_s\right)/d\right] = 25$ unknowns. Accordingly, the measured profile shows 25 height samples $B_j$ , $j$ = 0,1…25, as it is required by the system matrices in (A.29). Then, the Matlab® script computes the 26 x 26 = 676 matrix elements using the convolution theorem expression shown in equation (10). The solution of the system of 26 linear equations into 26 unknowns is performed by the Matlab® standard library. The total computation time, including the measured profile upload and the sample data capture, is of the order of one second on a standard computer. The complete list of 26 coverage coefficients is listed in the inset of **Figure 7(b)** and it constitutes the coverage spectroscopy of the measured CNT film. It is evident that the substrate surface is almost fully covered by CNTs, indicated by the coefficient $c_0$ = 4.15% corresponding to the uncovered $SiO_2$. The single-layer CNT coverage results $c_1$ = 24.1%. The largest coverage configuration belongs to the dual-layer CNT, represented by the coverage value $c_2$ = 35.95%. Other coverages report $c_3$ = 20.96% and $c_4$ = 8.48%, respectively for the triple and quadruple layers. The sensitivity of the coverage spectroscopy is proven by the accuracy of the calculations of the coverage coefficients belonging to the long tail of the measured profile. The total coverage corresponds therefore to the sum of all coefficients except $c_0$ and it gives $C$ = 95.85% with the unitary normalization factor. The blue curve visible in **Figure 7(b)** is obtained by superposing the 26 partial densities, each one weighted with the appropriate coverage coefficient. It is apparent that the computed coverage spectral decomposition (blue line) provides an extremely good fit of the measured curve (red line), confirming the validity of the proposed method.

We remark that the bell-shaped dashed curves shown in **Figure 7(b)** represent the partial height density function components of the corresponding CNT configuration and they are not Gaussian, even if the substrate profile has been modelled with the Gaussian density. In fact, each partial density is obtained from the multiple convolution of the deterministic diameter height density function, shown in **Figure 3**, with the substrate Gaussian density profile.

**Figure 8** shows the topography (a) and the height density (b) of a second sample of (7,5) CNTs wrapped with PFO deposited onto a $SiO_2$ substrate and measured with AFM in tapping-mode operation.

(a)

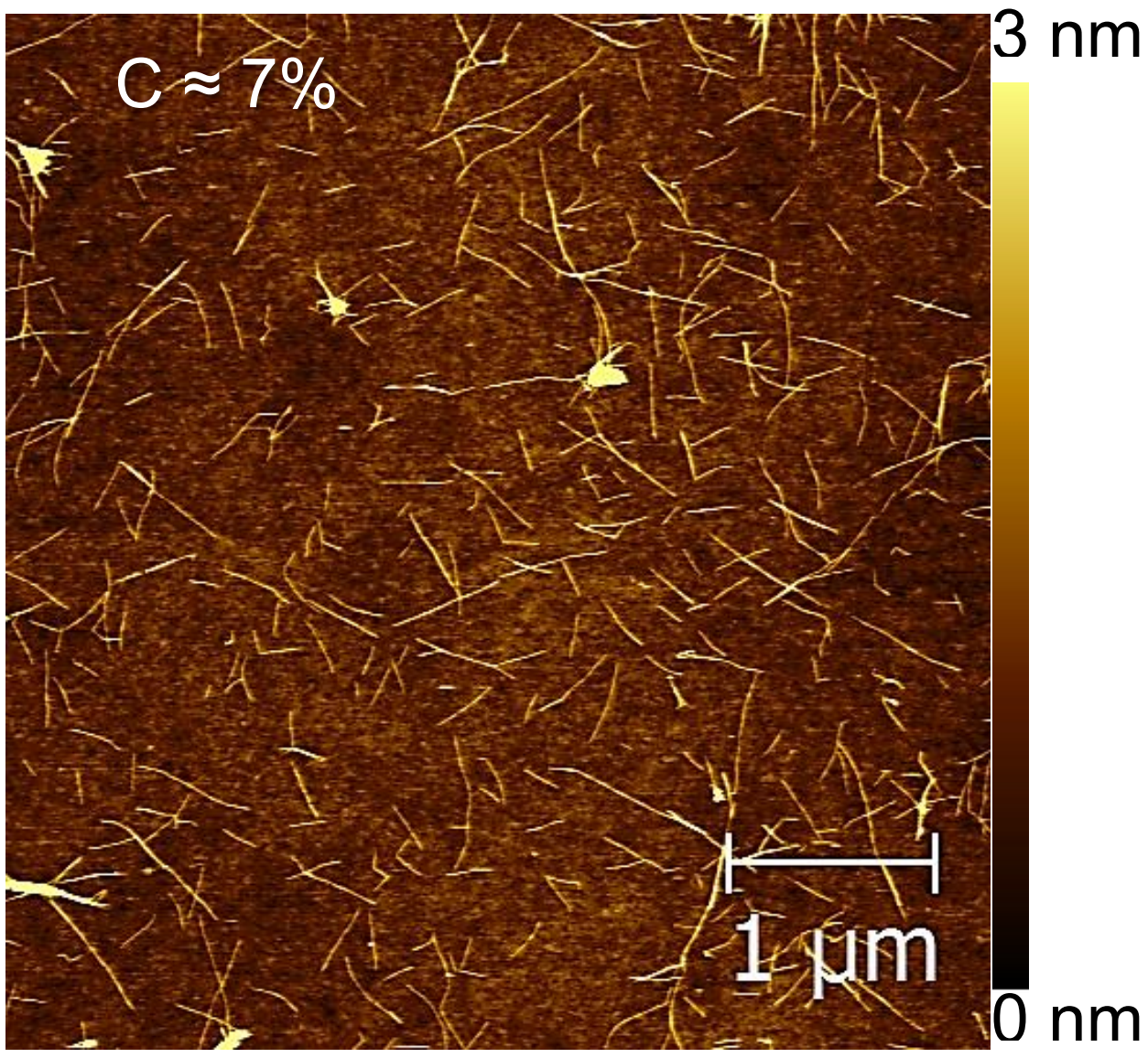

(b)

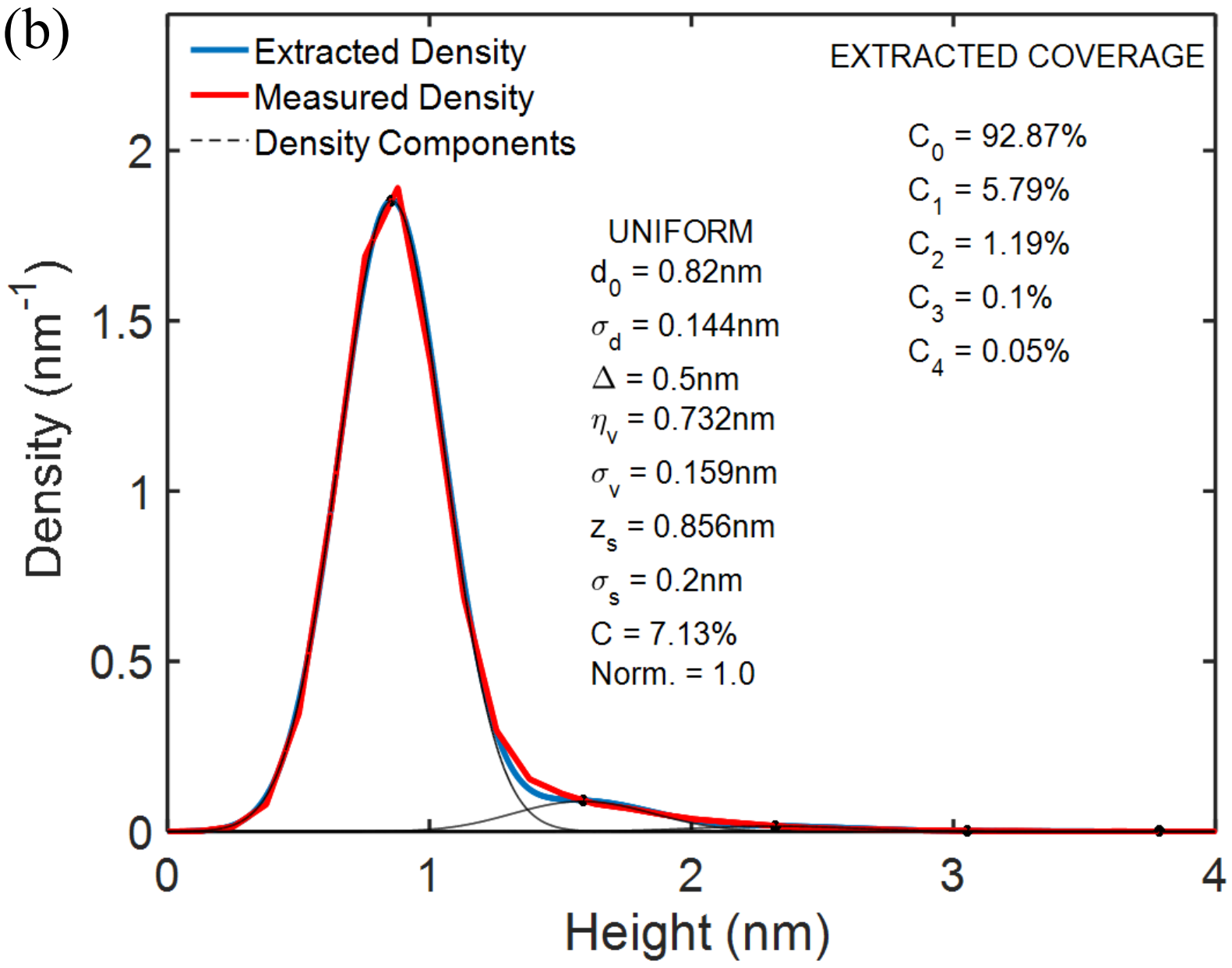

**Figure 8:** (a) Measured topography and (b) height density profile obtained with the AFM operated in tapping-mode on the solution processed PFO/(7,5) CNT network randomly distributed over the $SiO_2$ substrate. The coverage spectroscopy solution assumes the uniform diameter model and it is shown in the inset of (b). The computed total coverage results $C$ = 7.13%, with dominant single-layer CNT configurations with the coverage coefficient $c_1$ = 5.79%.

In this case, the CNT density is significantly lower than the sample shown in **Figure 7**, as most of the substrate area is well visible. Due to the much smaller number of stacking layers, the height density extinguishes faster, reaching the negligible tail contribution below the maximum height $z_{max}$ = 4 nm. The small percentage of covered area is confirmed by the relatively large peak of the substrate height density shown in **Figure 8(b)**. In this case, the CNT diameter has been modelled using the uniform density with the mean value $d$ = 0.82 nm[33] and the full width Δ = 0.144nm. The uniform diameter model is justified to account for additional polymer partially wrapped around some CNTs, making their effective diameter a statistical variable with uniform distribution. The choice of the diameter statistical model, i.e. deterministic, uniform or Gaussian-Harmonic, depends on the result of the fitting procedure. Some measurements fit better assuming the simpler diameter deterministic model, assuming that CNT population is almost free from any residual polymer. Both uniform and Gaussian-Harmonic diameter distributions represent either residual polymer contamination or heterogeneous CNT population with multiple diameters.

The substrate height density extracted from the measurement has a mean $\bar{z}_s$ = 0.856 nm and a standard deviation $\sigma_s$ = 0.2 nm. Because of the small value of the maximum height measured $z_{\max}$ = 4 nm, the number of allowed coverage coefficients is small as well, including only $N = \mathrm{int}\left[\left(z_{\max} - \bar{z}_s\right)/d\right]$ = 4 unknowns. The measured profile reports 4 height samples $B_j$ , $j$ = 0,…4, according to the system matrices shown in (A.29). The large percentage of uncovered substrate area is indicated by the coverage coefficient $c_0$ = 92.87%. The CNTs are almost distributed in the single-layer as indicated by the coverage coefficient $c_1$ = 5.79%. The dual, triple, and quadruple layer configurations give respectively $c_2$ = 1.19%, $c_3$ = 0.1%, and $c_4$ = 0.05%. The total coverage results therefore $C$ = 7.13% with the unitary normalization factor. Except for the small region between the tails of the substrate density and the partial density of the single-layer, the extracted curve (blue line) fits very well the curve of the measured profile (red line).

### 3.2 Silver Nanowires

In this section, we consider the AFM measurements of two samples of silver nanowires (Ag NW) deposited respectively from solution on a glass substrate and on a $SiO_2$ substrate. Silver nanowires are deposited on the substrate without any additional polymer and the measured height density is determined by the variation of the diameter along the nanowire itself, the different stacked superposition and the substrate roughness. However, silver nanowires have usually a much larger diameter than the substrate roughness, even for bare glass substrates, producing high resolution height measurements. The manufacturing process generates relatively large tolerances of the diameter among the Ag NW population, requiring either the uniform or the Gaussian-Harmonic diameter statistic to correctly represent the experimental evidence.

The inset of **Figure 9(a)** shows the topography of a single Ag NW located in a small area of the substrate. Despite the area scanned, the entire substrate was covered with many randomly distributed Ag NWs from the same batch with a nominal diameter $d$ = 75nm. Since the area measured by the AFM probe shows only one Ag NW sample, the statistical model of the diameter applies to the radial uniformity of the cylindrical nanostructure instead of the ensemble model of the entire Ag NW population. The height profile of the single Ag NW topography shown in the inset of **Figure 9(a)** was processed, and the height density profile shown in **Figure 9(b)** was obtained. In this case, the Gaussian-Harmonic diameter model was used, as suggested from the diameter distribution provided by the Ag NW manufacturer. The maximum height of the measured interval results $z_{max}$ = 120 nm. In order to model the diameter with the Gaussian-Harmonic density, a good match with the measured profile was found by setting the nominal diameter to $d$ = 75.7 nm with the standard deviation $\sigma_d$ = 0.459 nm. The glass substrate was modelled using the Gaussian density with mean height $z_s$ = 5.8 nm and standard deviation $\sigma_s$ = 2.4 nm. The simulated curves shown in **Figure 9(a)** are the density function of the Gaussian-Harmonic diameter (black line), the Gaussian substrate height (red line), the stand-alone Ag NW height (blue line) and the total height of the Ag NW placed onto the rough glass substrate (purple line). All these curves have unit area as they are probability densities. **Figure 9(b)** shows the measured height density (red line) and the superposed extracted height density (blue line) obtained with the coverage spectroscopy method. Since there is only one sample in the scanned area, the system of equations (9) contains only two linear equations with two unknown coverage coefficients, namely the uncovered substrate coefficient $c_0$ and the single layer coverage coefficient $c_1$. The inset in **Figure 9(b)** shows the detailed profiles of the measured and computed height densities. The computed curve provides a good fit to the measured profile and it also highlights the asymmetric shape of the peak, which is in agreement with the prediction of the theoretical model. The total coverage results $C = c_1$ = 5.48%.

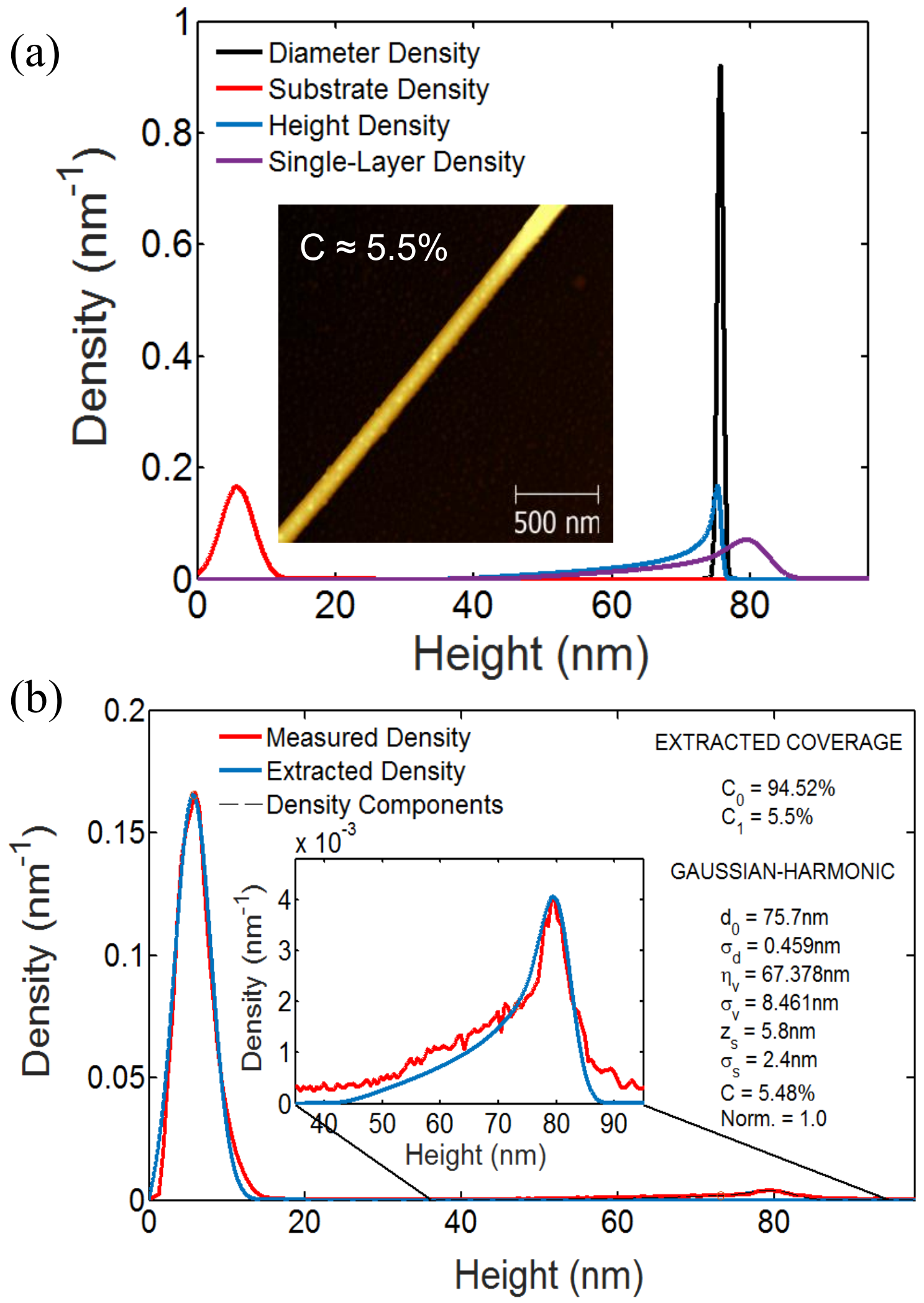

**Figure 9:** (a) Simulation of the density function of the Gaussian-Harmonic diameter (black line), the Gaussian substrate height (red line), the stand-alone Ag NW height (blue line) and the total height of the Ag NW placed onto the rough glass substrate (purple line). Inset: measured topography. (b) Height density profile obtained with the AFM operated in tapping-mode on the random network of silver nanowires distributed over the glass substrate. The solution of the coverage spectroscopy algorithm assumes the Gaussian-Harmonic diameter model and it is shown in the inset of (b): the total coverage results $C = c_1 = 5.48\%$ and corresponds to the coverage of the single Ag NW deposited in the scanned area.

The inset of **Figure 10(a)** shows the topography and the height density of a different sample of Ag NW deposited onto $SiO_2$ substrate. The diameter is larger as indicated by the right shift of the peak position in the height density function. The $SiO_2$ surface is flatter than the glass surface, and this is evident comparing the different Gaussian peak amplitudes of the substrate height densities shown in **Figure 9(b)** and **Figure 10(b)**.

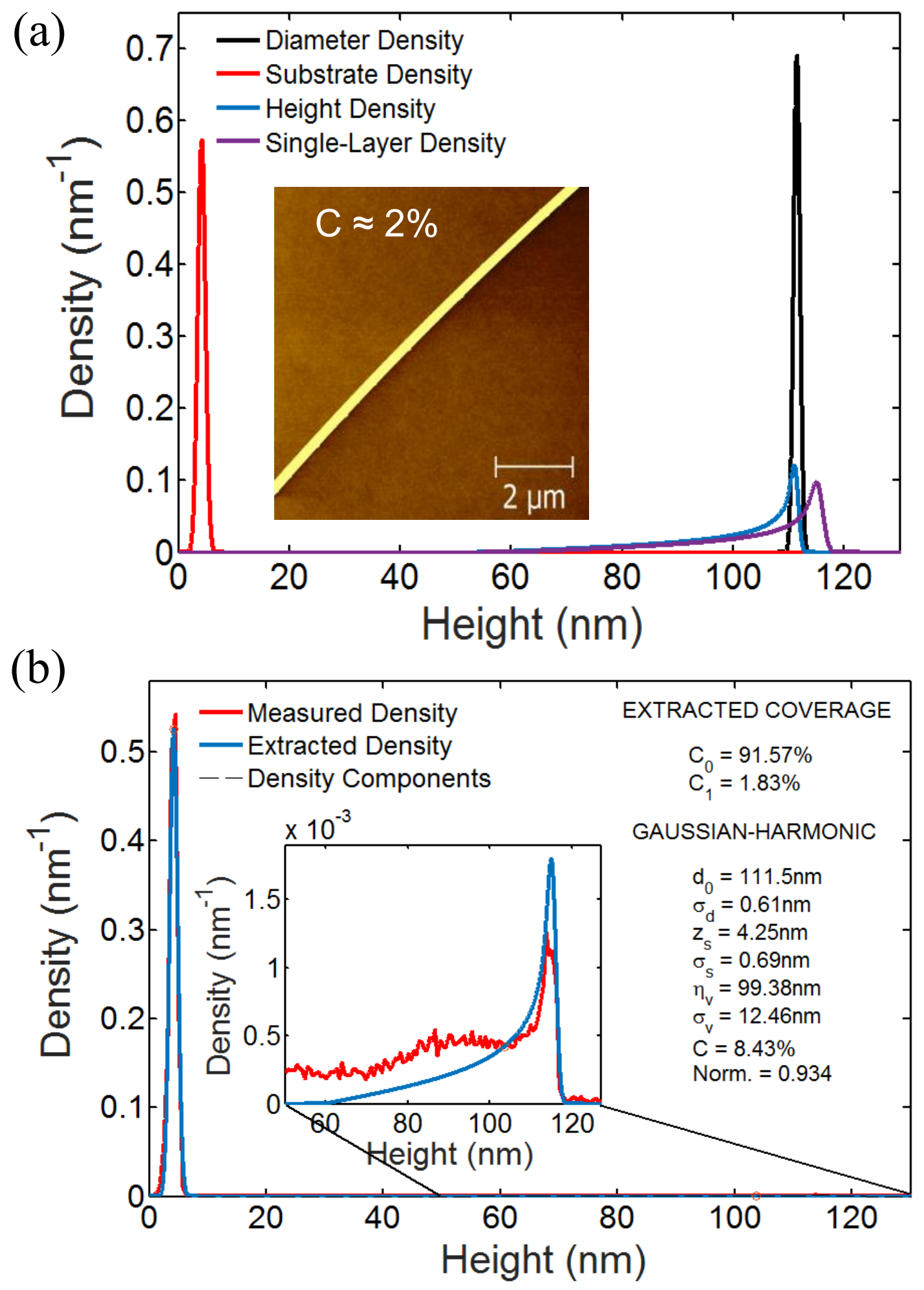

**Figure 10:** (a) Simulation of the density function of the Gaussian-Harmonic diameter (black line), the Gaussian substrate height (red line), the stand-alone Ag NW height (blue line) and the total height of the Ag NW placed onto the $SiO_2$ substrate (purple line). Inset: measured topography. (b) Height density profile obtained with the AFM operated in tapping-mode on the random network of silver nanowires distributed over the $SiO_2$ substrate. The solution of the coverage spectroscopy algorithm assumes the Gaussian-Harmonic diameter model and it is shown in the inset of (b): the total coverage corrected by the normalization factor results $C = c_1 = 1.96\%$ and corresponds to the coverage of the single Ag NW deposited in the scanned area. The uncorrected coverage instead is $C = 1.83\%$.

A good match for the measured profile was found by adjusting the nominal diameter to $d = 111.5$ nm with the standard deviation $\sigma_d = 0.61$ nm. The $SiO_2$ substrate was modelled using the Gaussian density with mean height $z_s = 4.25$ nm and standard deviation $\sigma_s = 0.69$ nm. The position of the height peak is well captured by the fitting model even if some residual fluctuations of the height density background, reasonably attributed to the measurement calibration, are still visible. In this case, the flat substrate emphasizes the sharp falling edge of the density profile at the nominal diameter value, confirming the theoretical model behavior we obtain when the diameter is much larger than the substrate roughness. Because of the low value reached by the normalization factor $\mathrm{N}$, the coverage coefficients must be corrected in this case, by dividing them by the normalization factor, i.e. $C = c_1/\mathrm{N} = 0.0183/0.934 = 0.0196$, corresponding to the uncovered area of the scanned surface equal to $C_0 = c_0/\mathrm{N} = 0.9157/0.934 = 0.9804$. We remark that these calculations consider the correction of the normalization factor $\mathrm{N} = 0.934$. In conclusion, the corrected total coverage is $C = 1.96\%$, while $C_0 = 98.04\%$.

At this point it is important to note that all scanning probe techniques present some native distortion when measuring highly resolved vertical profiles of isolated nanoparticles, like cylindrical nanoparticles with large diameters placed upon very low roughness surfaces. This is due to different probe resolutions available along vertical and lateral axes that distort the profile image and generate coverage coefficients larger than the expected/actual.

Unless the shape of the probe is known and de-convolved from the acquired data, the measured height profile will result distorted, mainly in the transversal direction, showing an artificial elliptical section instead of the expected circular one. The different native resolution available along vertical and lateral axes is determined by the different atomic force interaction established between the probe shape and the sample surface.

## Conclusions

In conclusion, we presented a new method for the calculation of the coverage coefficients of randomly distributed cylindrical nanoparticles, and their random networks, by using the topography obtained through AFM measurements. The diameter of CNTs, Ag NWs, and more generally of any cylindrical nanostructure, has been modelled as a random variable distributed with deterministic, uniform or Gaussian-Harmonic density function. The height density function has been derived for each diameter distribution and for any cylindrical aggregate order and it was used to generate the master coverage equation. The coverage spectroscopy method has been successfully tested on several aggregates of randomly distributed CNTs and Ag NWs, thus providing a functional and extremely useful new technique for a more accurate and in-depth surface characterization.

## Acknowledgements

The authors would like to acknowledge Professor T. Hertel and Ms. S. Fuchs from Julius-Maximilian University Wurzburg for the supply of polymer-wrapped carbon nanotubes. The research leading to these results has received funding from the People Program (Marie Curie Actions) of the European Union's Seventh Framework Program FP7/2007-2013/: “Polymer-carbon nanotubes active systems for photovoltaics” (POCAONTAS, under REA grant agreement n° 316633).

## Appendix A

*Material Preparation:* Single-walled carbon nanotubes (Sigma-Aldrich, CoMoCat SWNTs, diameter 0.7 – 0.9 nm, SWNT content ≥ 77%) were dispersed in toluene together with polyfluorene [poly(9,9-di-n-octylfluorenyl-2,7-diyl)] (Sigma-Aldrich, $M_w \geq 20000$). The mixture was then subjected to ultrasonication, centrifugation and vacuum filtration. A detailed description of the full procedure can be found in the work of Bottacchi et *al*.[33] Silver nanowires (Blue Nano, average diameter: 90 (±20) nm, average length: 30 μm) were dispersed in isopropyl alcohol in a concentration of 2 mg/ml. Both solutions were spin-coated at 1000 rpm for 30 s, followed by 15 min thermal annealing at 90°C to remove residual solvent.

*Surface Characterization:* Surface topography images and height distributions were obtained using the Agilent 5500 SPM atomic force microscope operating in tapping-mode. Budget Sensors Tap300Al-G silicon probes were used, with a spring constant of 40 $Nm^{-1}$ and a probe-radius <10 nm. Image planarization and simulations were performed using Gwyddion 2.38 and Matlab R2015a.

## Appendix B

$$f_{\underline{v}}^{(1)}(z) = \delta(z-d) \tag{.1}$$

$$\underline{v}^{(k)} = \sum_{j=1}^{k} \underline{v}_j \ , \quad f_{\underline{v}}^{(k)}(z) = \delta(z-d) * \delta(z-d) \mathrm{K} * \delta(z-d) = \delta(z-kd) \tag{.2}$$

$$f_{\underline{v}}(z) = \sum_{k=0}^{N} c_k f_{\underline{v}}^{(k)}(z) = \sum_{k=0}^{N} c_k \delta(z-kd) \tag{.3}$$

$$\sum_{k=0}^{N} c_k = 1 \qquad ( .4)$$

$$C_N = \sum_{k=1}^{N} c_k = 1 - c_0 \qquad ( .5)$$

$$f_{\underline{z}}(z) = f_{\underline{v}}(z) * g_{\underline{u}}(z) \qquad ( .6)$$

$$\sum_{k=0}^{N} c_k g_{\underline{u}}\left[z_s + (j-k)d\right] = B_j = f_{\underline{z}}\left(z_j\right), \quad z_j = z_s + jd, \quad j = 0,1 \ldots N \qquad ( .7)$$

$$\mathbf{A} = \begin{bmatrix} a_{00} & a_{01} & \cdots & a_{0N} \\ a_{11} & a_{12} & \cdots & a_{1N} \\ \cdots & \cdots & \cdots & \cdots \\ a_{N1} & a_{N2} & \ldots & a_{NN} \end{bmatrix}, a_{jk} = g_{\underline{u}}\left[z_s + (j-k)d\right]$$

$$\mathbf{c} = \begin{bmatrix} c_0 \\ c_1 \\ \ldots \\ c_N \end{bmatrix} \quad , \quad \mathbf{B} = \begin{bmatrix} B_0 \\ B_1 \\ \ldots \\ B_N \end{bmatrix} \qquad ( .8)$$

$$\mathbf{Ac} = \mathbf{B} \quad \Rightarrow \quad \mathbf{c} = \mathbf{A}^{-1}\mathbf{B} \qquad ( .9)$$

$$a_{jk} = \frac{e^{-\frac{(j-k)^2 d^2}{2\sigma_s^2}}}{\sigma_s\sqrt{2\pi}}, \quad j,k = 0,1 \ldots N \qquad ( .10)$$

$$f_{\underline{x}|\underline{y}}(x) = \begin{cases} 1/y\,, & |x| \le y/2 \\ 0 \quad, & elsewhere \end{cases} \qquad ( .11)$$

$$f_{\underline{xy}}(x,y) = f_{\underline{x}|\underline{y}}(x)\, f_{\underline{y}}(y) = \begin{cases} \dfrac{1}{y} f_{\underline{y}}(y), & |x| \le y/2 \\ 0 & , elsewhere \end{cases} \qquad ( .12)$$

$$\int_{-\infty}^{+\infty} f_{\underline{v}}(z)\,dz = \int_{-\infty}^{+\infty} \frac{2z - y}{y\sqrt{yz - z^2}}\,dz = 1 \qquad ( .13)$$

$$\eta_{\underline{v}} = \frac{d}{2}\left(1 + \frac{\pi}{4}\right) \approx 0.893\,d \qquad ( .14)$$

$$\sigma_{\underline{v}} = \frac{d}{2}\sqrt{\frac{2}{3} - \frac{\pi^2}{16}} \qquad ( .15)$$

$$f_{\underline{y}}(y) = \begin{cases} 1/\Delta\,, & |y-d| < \Delta/2 \\ 0, & elsewhere \end{cases} \qquad ( .16)$$

$$f_{\underline{a}}(\xi) = \frac{1}{\sigma\sqrt{2\pi}} e^{-\frac{(\xi-\rho)^2}{2\sigma^2}} \quad , \quad f_{\underline{b}}(\eta) = \frac{1}{\sigma\sqrt{2\pi}} e^{-\frac{(\eta-\rho)^2}{2\sigma^2}} \qquad ( .17)$$

$$\underline{y} = 2\sqrt{\underline{a}^2 + \underline{b}^2} \qquad ( .18)$$

$$yd >> 4\sigma^2 \quad \Rightarrow \quad f_{\underline{y}}(y) \approx \frac{e^{-\frac{(y-d)^2}{8\sigma^2}}}{2\sigma\sqrt{2\pi}} \quad \Rightarrow \quad \begin{cases} \eta_{\underline{y}} = d \\ \sigma_{\underline{y}} = 2\sigma \end{cases} \qquad ( .19)$$

$$\underline{\hat{z}}^{(k)} = \sum_{i=1}^{k} \underline{z}_i \qquad ( .20)$$

$$f_{\underline{v}|k}(z) = f_{\underline{v}}(z) * f_{\underline{v}}(z) * \cdots\ k\ times\,, \quad k = 1, 2 \ldots\,, \quad f_{\underline{v}|0}(z) = \delta(z)\,, \quad f_{\underline{v}|1}(z) = f_{\underline{v}}(z) \qquad ( .21)$$

$$\underline{z}^{(k)} = \underline{z}_s + \sum_{i=1}^{k} \underline{z}_i \qquad ( .22)$$

$$f_{\underline{z}|k}(z) = g_{\underline{u}}(z) * f_{\underline{v}|k}(z)\,, \quad f_{\underline{z}|0}(z) = g_{\underline{u}}(z) \qquad ( .23)$$

$$\sum_{k=0}^{N} c_k = 1 \qquad ( .24)$$

$$C = \sum_{k=1}^{N} c_k = 1 - c_0 \qquad ( .25)$$

$$f_{\underline{v}}(z) \overset{\Im}{\longleftrightarrow} F_{\underline{v}}(\xi) \quad \Rightarrow \quad \begin{cases} f_{\underline{v}|k}(z) \overset{\Im}{\longleftrightarrow} F_{\underline{v}}^{k}(\xi), \ k = 1, 2 \ldots N \\ \delta(z) = f_{\underline{v}|0}(z) \overset{\Im}{\longleftrightarrow} F_{\underline{v}}^{0}(\xi) = 1, \ k = 0 \end{cases} \qquad ( .26)$$

$$z_j = z_s + jd\,, \quad z_0 = z_s\,, \quad z_1 = z_s + d \ldots\,, \quad j = 0, 1 \ldots N \qquad ( .27)$$

$$a_{jk} = f_{\underline{z}|k}\left(z_s + jd\right) = \left[g_{\underline{u}}\left(z\right) * f_{\underline{v}|k}\left(z\right)\right]_{z=\overline{z}_s+jd}, \quad a_{j0} = g_{\underline{u}}\left(z_s + jd\right), \quad j,k = 0,\ldots N \tag{.28}$$

$$\mathbf{A} = \begin{bmatrix} a_{00} & a_{01} & \ldots & a_{0k} & \ldots & a_{0N} \\ a_{10} & a_{11} & \ldots & a_{1k} & \ldots & a_{1N} \\ \ldots & \ldots & \ldots & \ldots & \ldots & \ldots \\ a_{j0} & a_{j1} & \ldots & a_{jk} & \ldots & a_{jN} \\ \ldots & \ldots & \ldots & \ldots & \ldots & \ldots \\ a_{N0} & a_{N1} & \ldots & a_{Nk} & \ldots & a_{NN} \end{bmatrix}, \quad \mathbf{c} = \begin{bmatrix} c_0 \\ c_1 \\ c_2 \\ \ldots \\ c_k \\ \ldots \\ c_N \end{bmatrix}, \quad \mathbf{B} = \begin{bmatrix} B_0 \\ B_1 \\ B_2 \\ \ldots \\ B_k \\ \ldots \\ B_N \end{bmatrix} \tag{.29}$$